\documentclass{article}
\usepackage{amsmath,amsfonts,amssymb,graphicx,array,tabu,longtable,lineno,setspace,gensymb}
\title{Constants of explosive limits}
\author{Ali M. Nassimi$^a$\footnote{Corresponding author, Phone: 00989106615319, ali.nassimi@utoronto.ca}, Mohammad Jafari$^b$, Hossein Farrokhpour$^a$\\ and Mohammad H. Keshavarz$^b$\\
$^a$Department of Chemistry, Isfahan University of Technology,\\ Isfahan 8115683111, Iran\\
$^b$Department of Chemistry, Malek Ashtar University of\\ Technology, Shahin Shahr, 83145 115, Iran}

\begin{document}

\bibliographystyle{unsrt}
\maketitle

\begin{abstract}
 This work defines density factor as the ratio of before ignition density to after ignition density of the ignition mixture. This work provides an estimation method for explosive limits of various fuels under room temperature and pressure by showing that for a large universe of fuels, constant adiabatic flame temperature and density factor are appropriate approximations at the lower explosive limit while only a constant density factor might be an appropriate approximation at the upper explosive limit.
 Thus the assumption of a constant adiabatic flame temperature can be used in calculating lower explosive limit while the assumption of a constant density factor can be used in approximating upper explosive limit.

\end{abstract}

\textbf{List of abbreviations:}  Lower explosive limit (LEL),
upper explosive limit (UEL), explosive limit (EL),
adiabatic flame temperature (AFT), theoretical threshold temperature (TTT),
Equivalence ratio ($\phi$), Chemical Equilibrium with Applications (CEA),
Constant Enthalpy and Pressure (HP), Constant internal energy and volume (UV), Lower limit flame temperature (LLFT), Upper limit flame temperature (ULFT), Limit flame temperature (LFT), Lower limit density factor (LLDF), Upper limit density factor (ULDF), Limit density factor (LDF).

\section{Introduction}
Flammability parameters are important safety considerations in various applications ranging from the design of combustion chambers to the design of fuel tanks. Moreover, a knowledge of the flammability parameters is used in judging fire and explosion hazards of technological processes and technological installations.
Among these parameters are flash point~\cite{Kes,Kes3,Kes1}---the least temperature at which evaporation is enough for the vapor-air mixture above the flammable liquid to be ignitable---and explosive limits~\cite{bey}.
A mixture of a fuel and an oxidant can give rise to a self propagating flame if the oxidant/fuel ratio is within a specific range. The boundaries of this range are called explosive limits\footnote{Also called flammability limits.}.
 Lower explosive limit (LEL)---defined as the least fuel concentration capable of self-propagating a flame---and  upper explosive limit (UEL)---defined as the greatest fuel concentration capable of self-propagating a flame---are properties of a fuel-oxidizer mixture. Many studies measure LEL and UEL of various materials~\cite{627,Cash,680};
 in practice, most ignition processes involve a mixture of fuels and/or diluents, so the explosive limits (ELs) of fuel--fuel--oxidant (air) mixtures~\cite{Shoor,den,Miao,Shos} and fuel--oxidant--diluent mixtures \cite{sch,vidal} are studied. Further the effect of temperature \cite{den,rowl,gibbon1994,Kond,Mend} and pressure \cite{den,gibbon1994,Van,Arna} on ELs are studied.

 Both the LEL and the UEL depend on the temperature and pressure of the fuel-oxidizer mixture, but they also depend on the size, geometry and heat conductance of the combustion vessel and on the source of ignition \cite{Cowa,Raz,Gier}.
Measurement condition dependence causes values of ELs measured according to different standard methods to differ \cite{Sme,Zloch}. Common standard methods for measuring explosive limits are~\cite{Schro}: a) the US bureau of mines standard, which uses a vertical tube with an inner diameter of 5 cm and a length of 150 cm, with an electric spark or a pilot flame at the open lower end of the tube~\cite{Cowa}, b) the ASTM 681 standard, which uses a 5 dm$^3$ spherical glass with a central 15 kV igniter~\cite{ASTM}, c) the DIN 51 649 standard, which uses a cylindrical vertical glass with inner diameter of 6 cm, height of 30 cm and a 15 kV igniter located 6 cm above the tube's bottom~\cite{DIN51},
and d) the VDI 2263-1 standard developed for measuring dust explosion limits but also used for gaseous mixtures, which uses a 20 dm$^3$ spherical vessel with a central ignition~\cite{VDI22}.

The LEL and the UEL are generally considered to be 0.5 and 3 times the stoichiometric concentration of the fuel \cite{Tma}. Methods for calculating the LEL or the UEL make use of the standard enthalpy of combustion \cite{Suz,Brit}, normal burning velocity \cite{Mace}, adiabatic flame temperature (AFT) for partial combustion to CO and H$_2$O \cite{Shebe}, minimum spark ignition energy \cite{Fen}, ratio of stoichiometric AFT to AFT the EL~\cite{Mend} or constancy of AFT \cite{vidal,Mel,Shu}.
In calculation of flammability limits from AFT, the existence of the flammability limits is considered to be a consequence of existence of a minimum sustaining temperature for the flame---thermal theory \cite{Will}.
For LEL this sustaining temperature called theoretical threshold temperature (TTT) is claimed to be approximately a universal constant for every fuel \cite{vidal}. Thermal theory has also been extended to UEL \cite{Mash}. TTT at the upper and lower limits of explosivity should be distinguished but since we only use the concept and will not report any values for it, the same abbreviation is used for both temperatures.
 If we limit our attention to specific classes of compounds relations exist among flammability parameters, e.g., in the case of alkanes a linear relationship between molar heat of combustion and inverse of the LEL and another linear relationship between the LEL and the UEL exists \cite{Aff}.

To calculate explosive limits one needs an appropriate criteria for defining sustained ignition.
The most obvious ignition criterion is visual criterion, i.e., the flame moves away from the ignition source~\cite{EN1839}. Visual criterion is not useful for a theoretical study as the occurrence of observation of the flame propagation is difficult to quantify in a model.
Surrogates of the visual criterion are pressure rise criterion based on the rise of pressure by a fixed percentage of the initial pressure~\cite{EN1839} and temperature rise criterion based on a fixed amount of rise in the temperature of the fuel-oxidizer mixture \cite{Tsch}. The pressure rise criterion seems suitable for a study dealing with a constant internal energy and volume (UV) problem type. A UV problem type corresponds to an experimental setup where ignition occurs in a closed vessel.
The temperature rise criterion corresponds to the existence of the TTT and thus the constancy of the AFT at the ELs in a theoretical study based on thermodynamics. It can be used either for a UV or a constant enthalpy and pressure (HP) problem type.

Density factor (DF) is defined as the ratio of the density of the reactants mixture ($\rho_r$) to the density of the products mixture ($\rho_p$).
DF is introduced to serve as an indicator of ignition.
We assume equilibrium and ideal gas behavior. Mass conservation imply that DF is proportional to volume factor, i.e., ratio of after ignition to before ignition volume of the combustible mixture. Equilibrium and ideal gas assumptions imply that the volume factor is poportional to temperature factor, i.e., ratio of after ignition temperature to before ignition temperature of the combustible mixture. 
\begin{equation}
DF = \frac{\rho_r}{\rho_p} = \frac{\frac{m_r}{V_r}}{\frac{m_p}{V_p}} = \frac{V_p}{V_r} = \frac{\frac{n_pRT_p}{P_p}}{\frac{n_rRT_r}{P_r}} = \frac{n_p}{n_r}\frac{T_p}{T_r},
\end{equation}
where m, V, n, T and P respectively denote mass, volume, number of moles, temperature and pressure while indexes r and p respectively denote reactants and products.

In the following passages, we would use lower limit flame temperature (LLFT) for AFT at LEL, upper limit flame temperature (ULFT) for AFT at UEL and limit flame temperature (LFT) for AFT at either  EL. Further we use lower limit density factor (LLDF) for DF at the LEL, upper limit density factor (ULDF) for DF at the UEL and limit density factor (LDF) for DF at either EL.

Temperature in a flame is lower than the AFT by an amount which is proportional to the amount of heat loss per unit of the limiting reactant which in turn is proportional to surface/volume ratio of the combustion vessel.
LLFT and ULFT are upper bounds to the TTT's because AFT represents a situation of no heat loss from the flame.
In a first approximation TTT can be approximated with LFT but this approximation is not necessary; all we need is for the difference between LFT and TTT to be constant. LFT-TTT is proportional to the heat loss which under the same measurement conditions is proportional to the TTT.
Thus the constancy of this difference is implied by the constancy of the TTT.
Values as high as 1800 K to as low as 1000 K are suggested for TTT. This wide variation of TTT seems to be a result of variation in the combustion test conditions~\cite{Raz,Gier}.

Fuel and oxidizer contribute an amount of heat, $\Delta H_{com}$, together with an amount of quenching effect, $\int_{T_0}^{T_f}C_{p,mix}(T)dT$, where $T_0$ and $T_f$ are, respectively, the initial and flame temperature while $C_{p,mix}$ is the heat capacity of product mixture \cite{Bene}.
The AFT is derived by equating heat and quenching quantities \cite{Ma}.

This work considers flammables made up of the set of atoms \{C, H, N, O\}, among these flammables thermal theory fuel universe (TTFU) is defined as the set of flammables excluding Hydrogen (and deuterium), triple and adjacent double bond containing compounds and explosives.
For TTFU the values of LLFT and ULDF are shown to concentrate around their respective mean. This mean concentration is suggested as a method for predicting explosive limits when their experimental values are not available.

\section{Method} \label{CM}

Assuming a fuel molecule consisting of carbon, hydrogen, oxygen and nitrogen atoms; combustion equation can be written as
\begin{equation} \label{IgRe}
\phi C_iH_jO_kN_m + 4.773 l_s\text{ dry air} \rightarrow \text{combustion products},
\end{equation}
where, dry air = $0.7808N_2+0.2095O_2+0.0093Ar+0.0031CO_2$ (Thus molecular mass of dry air $M_{da}=29.0845$).
The stoichiometric oxygen requirement for each fuel molecule, $l_s=(2i+\frac{j}{2}-k)/2$. The equivalence ratio, $\phi$, is the ratio of fuel to oxygen over the stoichiometric ratio of fuel to oxygen. Equilibrium combustion products depend on the value of $\phi$.
In this analysis, we assume a premixed flame where products are determined by thermodynamic stability and not by diffusion rates.
 Using relations of mass action (chemical equilibrium constant relations) in finding equilibrium composition is equivalent to minimizing thermodynamic free energy of the mixture.

Chemical equilibrium with applications (CEA) \cite{Gord,McBr} calculates chemical equilibrium composition and properties of complex chemical mixtures.
CEA's algorithm uses Lagrange undetermined multipliers to minimize the relevant free energy and solves the resulting equations. Appropriate free energy to be minimized is determined by problem type.
Ignition in an open tube occurs at constant pressure and since ignition is a fast reaction it can be approximated as an adiabatic process. Thus the problem type in this work is HP.

\begin{figure}[!ht]
\centering
\includegraphics[width=.56\textwidth,angle=-90]{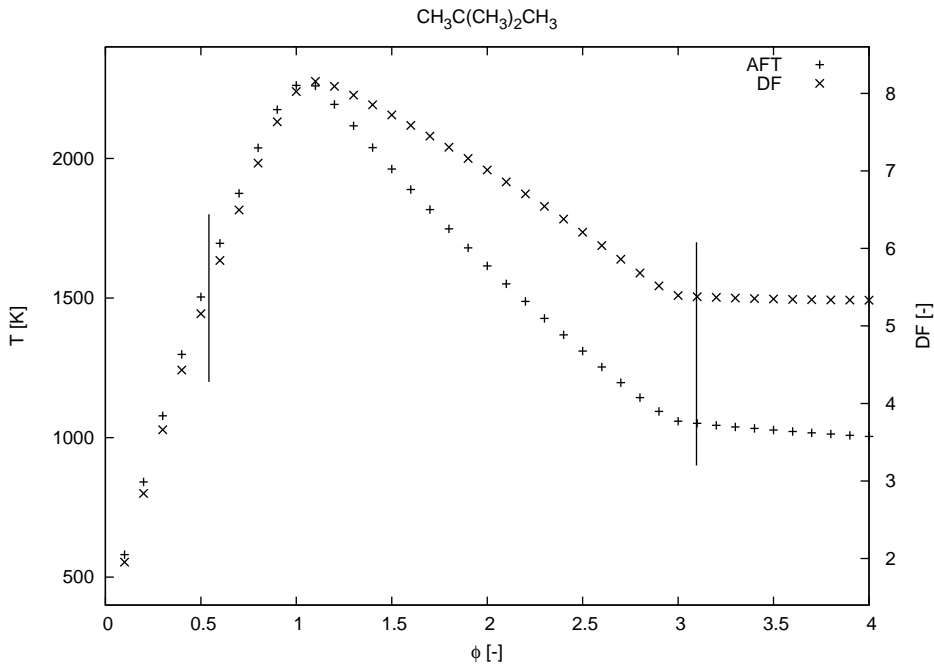}
\caption{Adiabatic flame temperature (left vertical axis) and density factor (right vertical axis) vs. equivalence ratio computed for 2,2-dimethylpropane.
The values of $\phi_{LEL}$ and $\phi_{UEL}$ are marked by vertical bars on the graph.} \label{pentane}
\end{figure}

\begin{figure}[!ht]
\centering
\includegraphics[width=0.56\textwidth,angle=-90]{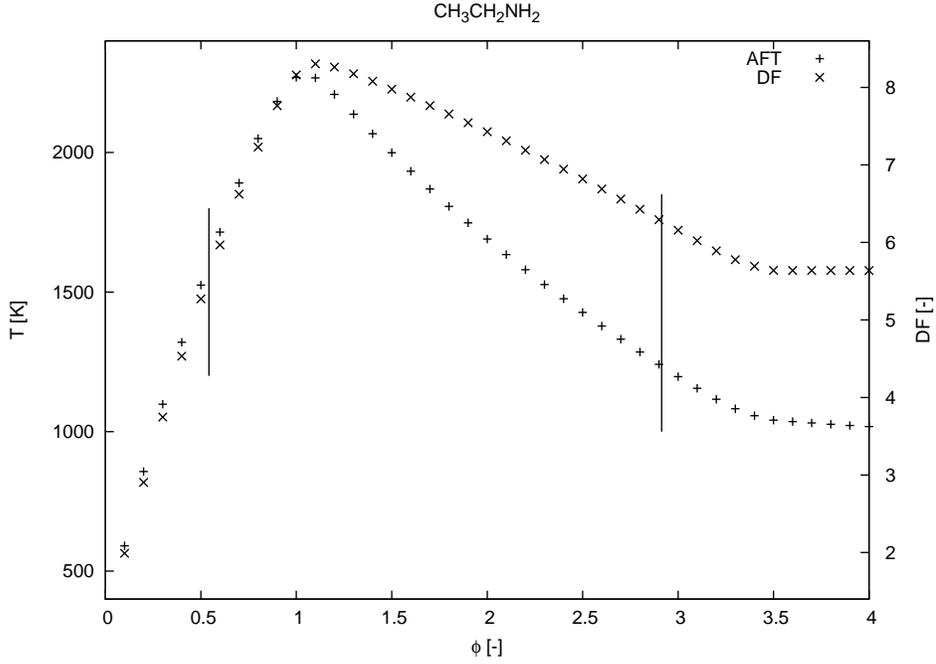}
\caption{Adiabatic flame temperature (left vertical axis) and density factor (right vertical axis) vs. equivalence ratio for ethylamine. The values of $\Phi_{LEL}$ and $\Phi_{UEL}$ are marked by vertical bars on the graph.} \label{propane}
\end{figure}

For each fuel, we perform chemical equilibrium calculations over a range of values of the equivalence ratio, $\phi=0.1,0.2,\cdots,4$ (except when it is necessary to go beyond this range to encompass the whole flammability range) and plot the AFT (DF) vs. $\phi$. Two examples are figures \ref{pentane} and  \ref{propane}.
 Reactant mixture density is calculated using the ideal gas law while the product mixture density is reported by CEA.
From the data presented in this graph we interpolate (derive with the assumption of linearity) values of the AFT (DF) corresponding to $\phi_{LEL}$ and  $\phi_{UEL}$, where  $\phi_{LEL}$ and  $\phi_{UEL}$, respectively, are the values of the equivalence ratio corresponding to the fuel concentration at the LEL and the UEL.

For compounds where experimental flammability limits are not available constancy of the LFT (LDF) may be used to derive explosive limits. To this end, one uses LFT (LDF) as a criterion for defining explosion. That is, one plots the AFT (DF) vs. $\phi$ to find the value of $\phi$ corresponding to the mean LFT (LDF) for the corresponding EL. One finds the volume percentage of fuel corresponding to the derived value of $\phi$ as the corresponding EL.

 Note the peak of both AFT and DF vs. $\phi$ graphs occur at values of equivalence ratio greater than 1 \cite{Law}.
Derivation of the AFT (DF) at ELs can be done by performing the chemical equilibrium calculation at the $\phi_{LEL}$ and $\phi_{UEL}$. But chemical equilibrium calculation cannot be done at assumed values of the LFT (LDF) to calculate the $\phi_{LEL}$ or $\phi_{UEL}$. To calculate the $\phi_{LEL}$ or $\phi_{UEL}$ we should draw a graph of the AFT (DF) vs. $\phi$. In order to keep the symmetry of deriving the LFT (LDF) from $\phi$ at the corresponding explosive limit and deriving $\phi$ at the corresponding EL from the LFT (LDF), we use the graph of AFT (DF) vs. $\phi$ for both calculations.

\section{Results}
For 52 fuels under study the values of LEL and UEL from \cite{627,Aff} are quoted in the second and third columns of table \ref{Data}. The LFTs (LDFs) are presented in the forth and fifth (sixth and seventh) columns of table \ref{Data}. Note that by neglecting few outliers the values for LFT (LDF) at either EL lay within a limited range.
The penultimate column of table \ref{Data} contains sum of mean deviation (SMD) for the LLFT and ULFT while the final column contains SMD for the LLDF and the ULDF. ELs are considered linear functions of temperature with very small slopes \cite{gibbon1994}, so small deviations from room temperature for few of the compounds reported in table \ref{Data} would not affect the results.

Table 2 reports the average values and standard deviations of the LFTs and LDFs. Standard deviation of the LLFT (LLDF) is about 15\% (16\%) of its mean value while the standard deviation of the ULFT (ULDF) is about 32\% (22\%) of its mean value.
Large ratio of standard deviation over the average value for the LFT (LDF) at both explosive limits  suggest that no single number can adequately represent the LFT (LDF) at either LEL or UEL.

{\centering
\begin{longtabu}{m{3.1cm}m{.95cm}m{1.05cm}m{1.15cm}m{1.15cm}m{1.15cm}m{1.15cm}m{1.25cm}m{1.15cm}}
\hline
Flammable gas&LEL vol.\% & UEL vol.\% & LLFT [K]&ULFT [K]&LLDF [-]&ULDF [-]&SMD LFT [K]&SMD LDF [-]\\ \hline
1,3-Butadiene&2.0&12.0&1671&1448&5.68&6.94&257&1.04 \\ 
1-Butene&1.6&10.0&1480&1211&5.06&6.05&171&0.47\\ 
2,2-Dimethylpropane&1.4&7.5&1585&1051&5.45&5.38&317&1.12\\ 
2-Butanone (Methyl Ethyl Ketone)&1.90&10.0&1539&1153&5.29&5.83&170&0.51\\ 
3-Methyl-1-Butene&1.50&9.1&1633&1142&5.58&5.89&275&0.75\\ 
Acetaldehyde diethyl acetal (Acetal)&1.6&10&1821&979&6.43&6.00&626&1.49\\ 
Acetone&2.6&13.0&1543&1201&5.33&5.93&126&0.45\\ 
Acetylene$^*$&2.5&100&1271&2848&4.23&10.40&1794&5.11\\ 
Acrylonitrile$^*$&3.0&17&1762&1772&5.94&7.97&672&2.32\\ 
Allene$^*$&1.5&11.5&1157&1891&3.89&8.04&951&3.09\\ 
Allyl alcohol (2-propen-1-ol)&2.5&18.0&1563&1081&5.36&6.05&265&0.36\\ 
Ammonia&15.0&28.0&1632&1823&5.70&6.93&593&1.04\\ 
Benzene (100 C)&1.3&7.9&1500&1411&5.09&6.41&127&0.28\\ 
Butane&1.8&8.4&1641&1147&5.67&5.71&277&1.01\\ 
Butyl Acetate (butyl ethanoate)&1.4&8.0&1588&1002&5.53&5.54&369&1.05\\ 
Carbon Monoxide$^*$&12.5&74&1391&1266&4.39&4.02&206&3.18\\ 
Cis-2-Butene&1.7&9.7&1539&1209&5.27&6.00&115&0.32\\ 
Cumene (isopropylbenzene)&0.9&6.5&1583&1183&5.32&5.91&183&0.45\\ 
Cyanogen$^*$&6.6&32&2236&2955&7.55&11.42&2329&7.38\\ 
Cyclohexane&1.3&7.8&1645&1062&5.69&5.54&367&1.20\\ 
Cyclopropane&2.4&10.4&1642&1598&5.60&7.16&378&1.18\\ 
Decane&0.7&5.0&1585&1033&5.40&5.61&335&0.84\\ 
Deuterium$^*$&4.9&75&708&1167&2.30&3.70&988&5.58\\ 
Dimethyl Ether&3.4&27.0&1567&974&5.42&5.81&376&0.66\\ 
Ethane&3.0&12.4&1535&1407&5.25&6.22&88&0.12\\ 
Ethanol&3.3&19.0 (60 C)&1493&1036&5.19&5.52&333&0.87\\ 
Ethyl Acetate (ethyl ethanoate)&2.2&11.0&1572&1095&5.42&5.72&260&0.75\\ 
Ethyl Benzene (100 C)&1.0&6.7&1560&1248&5.33&6.13&96&0.25\\ 
Ethylene&2.7&36&1372&1217&4.62&6.08&273&0.88\\ 
Ethylene Oxide (oxirane)$^*$&3.6&100&1553&1060&5.32&10.01&276&3.75\\ 
Heptane&1.05&6.7&1623&1033&5.64&5.60&374&1.09\\ 
Hexane&1.20&7.4&1610&1037&5.59&5.55&356&1.09\\ 
Hydrogen$^*$&4.0&75&628&1159&2.07&3.71&1075&5.80\\ 
Hydrogen Cyanide$^*$&5.6&40&1402&2235&4.66&8.47&1049&2.77\\ 
Isobutane&1.80&8.4&1638&1134&5.66&5.64&288&1.07\\ 
Isobutylene (2-methylpropene)&1.80&9.6&1596&1197&5.47&5.93&182&0.59\\ 
Methane&5.0&15.0&1463&1797&4.93&6.75&550&0.77\\ 
Methanol&6.7&36 (60 C)&1543&1042&5.33&5.48&285&0.90\\ 
Monoethylamine&3.5&14.0&1802&1235&6.29&6.28&351&1.06\\ 
Monomethylamine&4.9&20.7&1636&1361&5.67&6.63&135&0.71\\ 
n-Butanol&1.7&12.0&1507&991&5.21&5.54&363&0.83\\ 
Nonane$\dag$&0.80&6&1589&1016&5.48&5.67&356&0.87\\ 
Octane$\dag$&0.92&6.5&1618&1019&5.58&5.62&382&1.01\\ 
Pentane&1.40&7.8&1592&1055&5.51&5.49&320&1.07\\ 
Propane&2.1&9.5&1530&1331&5.26&6.18&17&0.14\\ 
Propene (Propylene)&2.4&11.0&1622&1442&5.53&6.61&202&0.56\\ 
Propylene Oxide (epoxypropane)&2.8&37&1682&1098&5.82&6.50&368&0.74\\ 
Propyne (Methylacetylene)$^*$&1.70&11.7&1257&1856&4.20&7.96&815&2.71\\ 
Styrene&1.10&6.1&1638&1455&5.58&6.75&231&0.75\\ 
Toluene (100 C)&1.20&7.1&1597&1316&5.44&6.23&65&0.27\\ 
Trans-2-Butene&1.7&9.7&1539&1203&5.23&5.93&120&0.42\\ 
Trimethylamine&2.0&12.0&1567&1104&5.42&6.07&246&0.40\\ 
\hline
\caption{Explosive limit data. LEL and UEL data are extracted from \cite{627}. These data relate to room temperature and atmospheric pressure except when another temperature is mentioned in the table. They are measured using a 2 inch tube with a spark ignition test setting. Values of LLFT and ULFT, (LLDF and ULDF) are derived using interpolation in the graph of the AFT (DF) vs. $\phi$ and expressed in Kelvin for LFTs and dimensionless units for LDFs. The penultimate (last) column contains sum of mean deviation for the LFTs (LDFs).
\newline $^*$ Excluded from TTFU
\newline $^\dag$ ELs from \cite{Aff}}
\label{Data}
\end{longtabu}}

At the UEL the assumption of chemical equilibrium does not apply very well as a large number of carbon containing compounds are formed among rich fuel combustion products while equilibrium calculations suggest CO, CO$_2$, CH$_4$ and C(s) as the only carbon containing products of combustion \cite{bey}.
This deviation from the equilibrium causes LFT-TTT to vary out of proportion to TTT (heat conduction is no more the only cause of LFT's deviation from TTT).
The greatest deviations from the mean LFT (LDF), as reported in the penultimate (last) column of table \ref{Data}, respectively, relate to cyanogen, acetylene, hydrogen, hydrogen cyanide, deuterium, allene, propyne, acrylonitrile,  and acetal
(cyanogen, hydrogen, deuterium, acetylene, ethylene oxide, carbon monoxide, allene and hydrogen cyanide).

It is suggested that the LFT and by implication the TTT are inversely proportional to reactivity of the fuel \cite{bey}. This is because the reactivity of a compound is determined by the activation energy of the rate determining step in units of $k_B T$ where $k_B$ is the Boltzmann constant.
Thus similar LFTs at an EL suggest close values of activation energy for the rate determining step of the respective fuel's ignition. Similar values of activation energy for the rate determining steps of a group of fuels ignition suggest similar rate determining steps for those reactions.

The group of fuels in table \ref{Data} which exhibit the greatest deviation from the mean LFT and LDF are compounds containing triple bond or adjacent double bounds, compounds capable of explosion in the absence of an oxidant\footnote{For acetylene and ethylene oxide UEL=100\%. For these compounds the AFT (DF) deviate greatly from the mean at the UEL.}, hydrogen and deuterium. Triple bond or adjacent double bond containing fuels are less stable than aromatic, mono-unsaturated and saturated fuels and one expects them to have an ignition mechanism considerably different from the later group of more stable fuels. Also hydrogen and deuterium have very high diffusivity; high diffusivity allows hydrogen and deuterium to sustain a flame at concentrations far lower than any other compound.

Aforementioned considerations suggest excluding triple bond and adjacent double bond containing compounds, hydrogen, deuterium and compounds capable of explosion; the remaining fuels would constitute an explosive limit thermal theory fuel universe (TTFU). TTFU constitute 81\% of all fuels in this study. Mean values and standard deviations for TTFU are presented in Table \ref{Aggregate}. The standard deviation of the LLFT is around 5\% of the mean LLFT. So one can consider a constant LLFT of 1586 as an acceptable approximation for the LLFT in TTFU. Standard deviation of the ULFT is about 17\% of the ULFT's mean value. Thus a constant ULFT is an inappropriate approximation.

\begin{table}
\centering
\begin{tabular}{p{2.cm}p{2.cm}lllll} \hline
&&LLFT &ULFT &LLDF &ULDF  \\ \hline
All fuels&Mean&1539&1323&5.27&6.32\\ 
&Std&232&422&0.83&1.40\\ 
&Std/Avg&0.151&0.319&0.158&0.222\\ 
&Min&628&974&2.07&3.70\\ 
&Max&2236&2955&7.55&11.42\\ 
&Range&1608&1981&5.47&7.72\\ 
TTFU&Ave&1586&1198&5.46&6.00\\ 
&Std&79&205&0.31&0.44\\ 
&Std/Avg&0.050&0.171&0.057&0.073\\ 
&Max&1821&1823&6.43&7.16\\ 
&Min&1372&974&4.62&5.38\\ 
&Range&449&848&1.81&1.79\\ 
\hline
\end{tabular}
\caption{Average values and standard deviations of adiabatic flame temperature and density factor at both explosive limits for all studied fuels and for the thermal theory fuel universe.}
\label{Aggregate}
\end{table}

LLDF and ULDF ratio of standard deviation to mean for TTFU of around 6\% and 7\%, respectively, suggest that the assumptions LLDF=5.46 and ULDF=6.0, in the TTFU are appropriate assumptions. We should note that ULDF (ULFT) for compounds with UEL=100\% are derived by calculation of equilibrium composition and not by interpolation in the DF (AFT) vs. $\phi$ graph.

\section{Conclusion}
The results presented in Table \ref{Aggregate} suggest that at the LEL the most aggregation around the mean occurs for the LFT in TTFU while at the UEL the most aggregation around the mean occurs for the LDF in the TTFU. Therefore, we suggest different methods for predicting each explosive limit. For the LEL we agree with the previous literature on using constancy of the LFT while noting the methods in-applicability for triple and adjacent double bond containing fuels and hydrogen. For the UEL we suggest using the LDF while noting the methods in-applicability for triple and adjacent double bond containing fuels, hydrogen and explosives.

Similar values of the LFT and the LDF in the TTFU suggest quite similar rate determining steps in the ignition reactions at the LEL for these compounds. Thus one can use the mean LFT or the mean LDF as a criterion for determining the LEL together with a graph similar to figure \ref{pentane} or \ref{propane} derived with CEA to calculate LEL for a mixture under specific conditions where an experimental value for the LEL is not available. For the UEL thermal theorie's prediction of a constant LFT for different fuels is inappropriate but the values of the LDF are more concentrated around their mean and one can use the LDF to derive an approximation to the value of the UEL using the graph of the DF vs. $\phi$.

With the assumptions of chemical equilibrium and ideal gas behavior constancy of each of the LFT and the LDF are equivalent. Yet in fuel rich mixtures the chemical equilibrium assumption does not hold very well and thus the concentration around the mean for the ULDF and the ULFT differ greatly.
The sample of compounds in Table \ref{Data} was chosen only based on ease of access to experimental data. The more important compounds for industrial applications likely have their safety data measured and published earlier than less important compounds. So we can claim compounds in Table \ref{Data} to form an importance sampling of flammable gases and vapors.
This work is related to experiments performed in an open ended reaction vessel. So we used a constant pressure assumption. In a future work we will use this method to calculate explosive limits.
\section*{Acknowledgment}
This work was supported by Iran's National Elite Foundation.

\bibliography{DMFTFL}

\begin{thebibliography}{10}

\bibitem{Kes}
M.~H. Keshavarz, M.~Jafari, M.~Kamalvand, A.~Karami, Z.~Keshavarz, A.~Zamani,
  and S.~Rajaee.
\newblock A simple and reliable method for prediction of flash point of
  alcohols based on their elemental composition and structural parameters.
\newblock {\em Process Saf Environ}, 102:1--8, 2016.

\bibitem{Kes3}
M.~H. Keshavarz and M.~Ghanbarzadeh.
\newblock Simple method for reliable predicting flash points of unsaturated
  hydrocarbons.
\newblock {\em J. Hazard. Mater.}, 193:335--41, 2011.

\bibitem{Kes1}
M.~H. Keshavarz, S.~Moradi, A.~R. Madram, H.~R. Pouretedal, K.~Esmailpour, and
  A.~Shokrolahi.
\newblock Reliable method for prediction of the flash point of various classes
  of amines on the basis of some molecular moieties for safety measures in
  industrial processes.
\newblock {\em J Loss Prevent Proc}, 26:650--9, 2013.

\bibitem{bey}
C.~Beyler.
\newblock Flammability limits of premixed and diffusion flames.
\newblock In M.J. Hurley, editor, {\em SFPE Handbook of Fire Protection
  Engineering}. Society of Fire Protection Engineers, 2016.

\bibitem{627}
M.~G. Zabetakis.
\newblock Flammability characteristics of combustible gases and vapors.
\newblock Technical Report Bulletin 627, US Bureau of mines, 1965.

\bibitem{Cash}
K.~Cashdollar, I~Zlochower, G.~Green, R.~Thomas, and M.~Hertzberg.
\newblock Flammability of methane, propane, and hydrogen gases.
\newblock {\em J Loss Prevent Proc}, 13:327--40, 2000.

\bibitem{680}
J.~Kuchta.
\newblock Investigation of fire and explosion accidents in the chemical,
  mining, and fuel-related industries---a manual.
\newblock Technical Report Bulletin 680, US Bureau of mines, 1985.

\bibitem{Shoor}
F.~{Van den Schoor}, R.~Hermanns, J.~van Oijen, F.~Verplaetsen, and L.~de~Goey.
\newblock Comparison and evaluation of methods for the determination of
  flammability limits, applied to methane/hydrogen/air mixtures.
\newblock {\em J Hazard Mater}, 150:573--81, 2008.

\bibitem{den}
F.~{Van den Schoor}, F.~Verplaetsen, and J.~Berghmans.
\newblock Calculation of the upper flammability limit of methane/hydrogen/air
  mixtures at elevated pressureschs and temperatures.
\newblock {\em Int J Hydrogen Energ}, 33:1399--406, 2008.

\bibitem{Miao}
H.~Miao, L.~Lu, and Z.~Huang.
\newblock Flammability limits of hydrogen-enriched natural gas.
\newblock {\em Int J Hydrogen Energ}, 36:6937--47, 2011.

\bibitem{Shos}
Y.~Shoshin and L.~de~Goey.
\newblock Experimental study of lean flammability limits of
  methane/hydrogen/air mixtures in tubes of different diameters.
\newblock {\em Exp. Therm Fluid Sci}, 34:373--80, 2010.

\bibitem{sch}
F.~{Van den Schoor}, F.~Norman, K.~Vandermeiren, F.~Verplaetsen, J.~Berghmans,
  and E.~Van den Bulck.
\newblock Flammability limits, limiting oxygen concentration and minimum inert
  gas/combustible ratio of {H$_2$/CO/N$_2$}/air mixtures.
\newblock {\em Int J Hydrogen Energ}, 34:2069--75, 2009.

\bibitem{vidal}
M.~Vidal, W.~Wong, W.J. Rogers, and M.S. Mannan.
\newblock Evaluation of lower flammability limits of fuel-air-diluent mixtures
  using calculated adiabatic flame temperatures.
\newblock {\em J Hazard Mater}, 130:21--7, 2006.

\bibitem{rowl}
J.R. Rowley, R.L. Rowley, and W.V. Wilding.
\newblock Estimation of the lower flammability limit of organic compounds as a
  function of temperature.
\newblock {\em J Hazard Mater}, 186:551--7, 2011.

\bibitem{gibbon1994}
HJ~Gibbon, J~Wainwright, and RL~Rogers.
\newblock Experimental determination of flammability limits of solvents at
  elevated temperatures and pressures.
\newblock In {\em Institution of Chemical Engineers Symposium Series}, volume
  134, pages 1--12. Hemisphere Publishing Corporation, 1994.

\bibitem{Kond}
S.~Kondo, K.~Takizawa, A.~Takahashi, and K.~Tokuhashi.
\newblock On the temperature dependence of flammability of gases.
\newblock {\em J Hazard Mater}, 187:585--90, 2011.

\bibitem{Mend}
A.~Mendiburu, J.~de~Carvalho~Jr, C.~Coronado, and G.~Chumpitaz.
\newblock Deteremination of lower flammability limits of {C-H-O} compounds in
  air and study of initial temperature dependence.
\newblock {\em Chem Eng Sci}, 144:188--200, 2016.

\bibitem{Van}
F.~{Van den Schoor} and F.~Verplaetsen.
\newblock The upper flammability limit of methane/hydrogen/air mixtures at
  elevated pressures and temperatures.
\newblock {\em Int J Hydrogen Energ}, 32:2548--52, 2007.

\bibitem{Arna}
J.~Arnaldos, J.~Kasal, and E.~Planas-Cuchi.
\newblock Prediction of flammability limits at reduced pressures.
\newblock {\em Chem Eng Sci}, 56:3829--3843, 2001.

\bibitem{Cowa}
H.~F. Coward and G.~W. Jones.
\newblock Limits of flammability of gases and vapors.
\newblock Technical Report Bulletin 503, US Bureau of mines, 1952.

\bibitem{Raz}
D.~Razus, M.~Molnarne, and O.~Fab.
\newblock Limiting oxygen concentration evaluation in flammable gaseous
  mixtures by means of calculated adiabatic flame temperatures.
\newblock {\em Chem Eng Process}, 43:775--84, 2004.

\bibitem{Gier}
M.~Gieras, R.~Klemens, A.~Kuhl, P.~Oleszczak, W.~Trzcinski, and P.~Wolanski.
\newblock Influence of the chamber volume on the upper explosion limit for
  hexane-air mixtures.
\newblock {\em J Loss Prevent Proc}, 21:423--36, 2008.

\bibitem{Sme}
G.~De Smedt, F.~de~Corte, and R.~Notele.
\newblock Comparison of two standard test methods for determining explosion
  limits of gases at atmospheric conditions.
\newblock {\em J Hazard Mater}, A70:105--13, 1999.

\bibitem{Zloch}
I.~Zlochower and G.~Green.
\newblock The limiting oxygen concentration and flammability limits of gases
  and gas mixtures.
\newblock {\em J Loss Prevent Proc}, 22:499--505, 2009.

\bibitem{Schro}
V.~Schroder and M.~Molnarne.
\newblock Flammability of gas mixtures part 1: fire potential.
\newblock {\em J Hazard Mater}, 121:37--44, 2005.

\bibitem{ASTM}
{ASTM E}681-09.
\newblock Standard test method for concentration limits of flammability of
  chemicals (vapors and gases), 2009.

\bibitem{DIN51}
{DIN 5}1649-1.
\newblock Determination of limits of flammability of gases and gas mixtures in
  air, 1985-2004.

\bibitem{VDI22}
{VDI 2263 Part I}.
\newblock Test methods for the determination of the safety characteristics of
  dusts, 1990.

\bibitem{Tma}
T.~Ma, Q.~Wang, and M.~Larranaga.
\newblock Correlations for estimating flammability limits of pure fuels and
  fuel-inert mixtures.
\newblock {\em Fire Saf J}, 56:9--19, 2013.

\bibitem{Suz}
T.~Suzuki.
\newblock Empirical relationship between lower flammability limits and standard
  enthalpies of combustion of organic compounds.
\newblock {\em Fire and Mater.}, 18:333--6, 1994.

\bibitem{Brit}
L.~Britton and D.~Frurip.
\newblock Further uses of the heat of oxidation in chemical hazard assessment.
\newblock {\em Process Saf Prog}, 22:1--19, 2003.

\bibitem{Mace}
A.~Macek.
\newblock Flammability limits: thermodynamics and kinetics.
\newblock Technical Report 76-1076, Natl. Bureau Standards of USA, 1976.

\bibitem{Shebe}
Yu.~N. Shebeko, W.~Fan, I.A. Bolodian, and V.~Yu. Navzenya.
\newblock An analytical evaluation of flammability limits of gaseous mixtures
  of combustible-oxidizer-diluent.
\newblock {\em Fire Saf. J.}, 37:549--68, 2002.

\bibitem{Fen}
J~Fenn.
\newblock Lean flammability limit and minimum spark ignition energy.
\newblock {\em Ind Eng Chem}, 43(12):2865--9, 1951.

\bibitem{Mel}
G.~Melhem.
\newblock A detailed method for estimating mixture flammability limits using
  chemical equilibrium.
\newblock {\em Process Saf. Prog.}, 16:203--18, 1997.

\bibitem{Shu}
G~Shu, B~Long, H~Tian, H~Wei, and X~Liang.
\newblock Evaluating upper flammability limit of low hydrocarbon diluted with
  an inert gas using threshold temperature.
\newblock {\em Chem Eng Sci}, 138:810--3, 2015.

\bibitem{Will}
FA~Williams.
\newblock {\em Combustion theory.}
\newblock Addison Wesley publishing company, London, 1969.

\bibitem{Mash}
C.~Mashuga and D.~Crowl.
\newblock Flammability zone prediction using calculated adiabatic flame
  temperatures.
\newblock {\em Process Saf. Prog.}, 18(3):127--34, 1999.

\bibitem{Aff}
W.~Affens.
\newblock Flammability properties of hydrocarbon fuels.
\newblock {\em J Chem Eng Data}, 11(2):197--202, 1966.

\bibitem{EN1839}
{EN }1839.
\newblock Determination of explosion limits of gases and vapours, 2012.

\bibitem{Tsch}
R~Tschirschwitz, V~Schr€oder, E~Brandes, and U~Krause.
\newblock Determination of explosion limits---criterion for ignition under
  nonatmospheric conditions.
\newblock {\em J Loss Prevent Proc}, 36:562--8, 2015.

\bibitem{Bene}
A.~{Di Benedetto}.
\newblock The thermal/thermodynamic theory of flammability: The adiabatic
  flammability limits.
\newblock {\em Chem Eng Sci}, 99:265--73, 2013.

\bibitem{Ma}
T~Ma.
\newblock A thermal theory for estimating the flammability limits of a mixture.
\newblock {\em Fire saf. J.}, 46:558--67, 2011.

\bibitem{Gord}
S.~Gordon and B.~J. McBride.
\newblock Computer program for calculation of complex chemical equilibrium
  compositions and applications {I. A}nalysis.
\newblock Technical Report 1311, NASA, 1994.

\bibitem{McBr}
B.~J. McBride and S.~Gordon.
\newblock Computer program for calculation of complex chemical equilibrium
  compositions and applications { II. U}sers manual and program description.
\newblock Technical Report 1311, NASA, 1996.

\bibitem{Law}
C.K. Law, A.~Makino, and T.F. Lu.
\newblock On the off-stoichiometric peaking of adiabatic flame temperature.
\newblock {\em Combust Flame}, 145:808--19, 2006.

\end{thebibliography}

\end{document}